\documentclass[conference]{IEEEtran}

\pdfoutput=1

\usepackage{cite}
\usepackage{amsmath,amssymb,amsfonts}
\usepackage{algorithmic}
\usepackage{graphicx}
\usepackage{textcomp}
\usepackage{xcolor}
\usepackage{url}
\usepackage{flushend}

\def\BibTeX{{\rm
B\kern-.05em{\sc i\kern-.025em b}\kern-.08em
T\kern-.1667em\lower.7ex\hbox{E}\kern-.125emX}}

\begin{document}

\title{The Dark Side of Unikernels for Machine Learning}

\author{
	\IEEEauthorblockN{Matthew Leon}
	\IEEEauthorblockA{
		\textit{Vanderbilt University}\\
		Nashville, Tennessee, USA\\
		matthew.leon@vanderbilt.edu
	}
}

\maketitle

\begin{abstract}
	This paper analyzes the shortcomings of unikernels as a method
	of deployment for machine learning inferencing applications as
	well as provides insights and analysis on future work in this
	space. The findings of this paper advocate for a tool to enable
	management of dependent libraries in a unikernel to enable a
	more ergonomic build process as well as take advantage of the
	inherent security and perfomance benefits of unikernels.
\end{abstract}


\section{Introduction}

Virtualization technology is used in datacenters spanning the
whole world to provide availability, scalability, and security to
millions of client workloads. While virtualizing an entire
computer and running everything, including the OS image,
libraries and application code, on top is still popular, many
alternative methods have emerged over the last decade, each with
promises to increase security and improve resource utilization.
There are numerous benefits to improving resource utilization of
the host system. For users, fewer resources used by the host
operating system means more resources available to application
code. For the large corporations hosting public clouds,
maximizing utilization on a small number of machines means lower
costs, as datacenters consume massive amounts of
power~\cite{directenergy}.  Recently, containerization technology
has been adopted as a way to reduce the number of virtualization
layers in the modern datacenter, with services like Docker
providing benefits including easier deployment, ensuring
consistency between the development and production environments,
providing limits on resource usage, and sandboxing applications
for better security.  Containerization cuts down on costs by
reducing duplication of the operating system—rather than running
several different stacks all virtualized on top of a hypervisor,
a user can run a single operating system and divide up its
resources among several containers. Unikernel is yet another
lightweight virtualization technology increasingly being adopted
in cloud data centers.

\section{What are Unikernels?}

Unikernels, on the other hand, focus on the other side of the
playing field from containers. With unikernels, the operating
system is totally eliminated—the application code itself is
augmented with the minimal set of code necessary to interface
with the hypervisor and is then directly run as a bootable image
on top of a hypervisor. The compactness of this system can result
in numerous benefits over containers and fully virtualized linux
servers. In one study, boot times as low as 50ms were achieved,
as well as lower memory usage and reduced latency due to
zero-copy network implementation~\cite{libos}. The significant
benefits of the unikernels are discussed in the next section.

\section{Study Goals}

This study aimed to analyze the state of a few different
unikernels and their environments, comparing them to traditional
methods of virtualization in terms of developer experience,
performance, flexibility, security, and feasibility for adoption.
Specifically, the study was conducted through a use case where we
wanted to understand whether it was feasible to deploy a machine
learning-trained image classification inference inside a
Unikernel. To that end, we implemented an image classification
API capable of receiving an image via HTTP and responding with an
inference as to the contents of the image.

\section{Report Organization}

This report first outlines preliminary knowledge about the
differences of unikernels, including major vendors of unikernel
technology as well as an overview of the pertinent differences
from ordinary virtualization solutions. The next section provides
an overview of the work done in the process of evaluating
the maturity of unikernels as a modern, lightweight alternative
to containerization technology. Finally, the paper is concluded
with an analysis of the hurdles that must be addressed before
unikernels are sufficient for a modern deployment,

\section{Unikernels In-depth}

\subsection{What are Unikernels?}

Unikernels, on the other hand, focus on the other side of the
playing field from containers.  With unikernels, the operating
system is totally eliminated—the application code itself is
augmented with the minimal set of code necessary to interface
with the hypervisor and is then directly run as a bootable image
on top of a hypervisor. The compactness of this system can result
in numerous benefits over containers and fully virtualized linux
servers. In one study, boot times as low as 50ms were achieved,
as well as lower memory usage and reduced latency due to
zero-copy network implementation~\cite{libos}. The significant
benefits of the unikernels are discussed in the next section.

Unikernels can be grouped into two distinct categories. Firstly
are unikernels that function as a library operating system. OSs
in this group, such as IncludeOS~\cite{includeos},
HaLVM~\cite{halvm}, and MirageOS~\cite{mirageos}, cannot run full
executable programs, instead, they are written in and run code in
an augmented runtime environment that implements operating system
functions, such as I/O. The other group of unikernels, such as
RumpRun~\cite{rumprun}, and Nanos~\cite{nanos}, provide
application binaries an entire POSIX-compatible runtime
environment which can run arbitrary ELF executables. In addition
to these runtime environments, several build, orchestration, and
packaging tools are available, such as ops~\cite{nanos},
Unikraft~\cite{unikraft}, and UniK~\cite{unik}. This study
investigates the feasibility and shortcomings of using these
tools to deploy a deep neural network inference solution
available via a web API.\@

\subsection{Benefits of Unikernels}

The single address space architecture of unikernels provides
numerous benefits that are not achievable with conventional
preemptive multitasking operating systems. Firstly, the total
attack surface is much lower with a unikernel. Bratterud, Happe,
and Duncan highlight a 92\% reduction in total bytes of code in a
running unikernel, which they translate to a 92\% smaller attack
surface~\cite{enhancingprivacy}. The lack of a shell prevents an
entire class of vulnerabilities, while a single address space
allows for compile-time address space layout randomization, which
is more performant than the runtime alternative. In addition to
the security implications of a single address space, the removal
of kernel space eliminates time spent in kernel space context
switches as well as scheduling interrupts by the guest OS.\@
Instead, scheduling and load balancing is handled entirely by the
hypervisor.

In terms of load balancing itself, unikernels offer distinct
benefits for web-related tasks, especially due to their startup
time. The unikernel itself being the executable and thus not
requiring file systems to be initialized as well as the small
size the kernel code occupies means that the only boot step
necessary is initializing the network interface. In a hypervisor
environment, this allows the unikernel to be booted in response
to an incoming request in time to handle that request. Such a
fast boot time allows horizontal scaling with the granularity of
individual requests. This instant availability enables
applications such as fog deployment for IoT, which was
investigated by Cozzolino, Ding, and Ott~\cite{fades}. This work
is further being applied at the same time as this research as
infrastructure in smart city monitoring of ongoing road
hazards~\cite{ecco}.

\section{Insights from our Study}

Supplementary source code materials and motivating examples for
the following findings may be found at~\cite{myghrepo}. Many
simple implementations of image classifications are available on
GitHub, such as~\cite{pytorchapi}. In the goal of evaluating the
effectiveness of unikernels in different environments and
implementations, three different machine learning frameworks were
tested: Tensorflow, PyTorch, and Tensorflow.js (Tensorflow and
Tensorflow.js are included separately as they do not share
bindings to the same underlying library; they are completely
separate implementations in two different languages of the same
API). IncludeOS was used in conjunction with Tensorflow, and
RumpRun and Nanos were both used to test each of PyTorch and
Tensorflow.js.

Our findings revealed that none of the tested solutions were
successful. The shortcomings ranged depending on the
implementation—Tensorflow and PyTorch struggled with issues
linking inside of the unikernel, and Tensorflow.js struggled
fetching the trained model via URL due to the lack of a DNS
resolver in the unikernel environment. When adding the node.js
extension to Tensorflow.js to allow for loading the model from
within the image, the unikernel struggled due to lack of node-gyp
(a C/C++ native binding) support inside the unikernel. We note
that Tensorflow.js could be extended to support loading from file
without involving node-gyp, but performing large modifications to
the source of the application was out of scope for this study’s
investigation of unikernels as an alternative deployment
environment. PyTorch encountered similar issues as it is an
optimized runtime with most of the deep learning code implemented
in C—the modules for the library were unable to be loaded inside
the unikernel environment.

Seeing as most of the encountered issues were due to the lack of
interoperability between native libraries and interpreted code,
the next approach we took was compiling Tensorflow into an
application compiled with IncludeOS, the library operating system
capable of transforming the C/C++ application it is built with to
an Xen-bootable executable.  Unfortunately, linking also became
an issue in this case. The publicly available distributions of
Tensorflow depend on over 10 shared libraries, and IncludeOS must
be built statically, which is not supported (nor possible in an
unsupported fashion) in any version of the library. Copying the
shared libraries into the image from the system used to build
Tensorflow resulted in a bootable system, but the execution
failed due to missing symbols in the outdated version of glibc
used in the host system. No other languages were tested after
these failures, as all languages link to the C library, with the
only exception being the previously mentioned Tensorflow.js
without node.js extensions, which is designed for the browser
environment. It was unexplored whether other smaller toolkits
would’ve been more successful—mlpack~\cite{mlpack} appears to be
a good candidate for future research, as it may allow static
linking~\cite{mlpackcmake}.

\section{Analysis and Possibilities for Future Work}

Unikernels, when compared to a deployment solution using docker
containers or a native Linux virtual machine, still have many
hurdles to overcome before they can claim full parity in terms of
supported use cases. Due to the decades of prevalence of
ecosystems which support dynamic linking as a way to quickly fix
security issues and reduce compiled code duplication across
binaries, even common libraries like Tensorflow do not support
static linking, which is unfortunate news for any application
developer looking to use these libraries in a unikernel. There
are ways to build a static library manually such as by packing
GCC’s object file output with tools such as ar, but these are
steps for build system maintainers rather than application
developers~\cite{ar}. It is this researcher’s opinion that
unikernels would be most benefited by a robust build tool which
handles dependency bundling inside the unikernel environment,
much like Docker’s \textit{build} command or Ansible scripts.
With access to a layered build system, unikernels could provide a
compelling base layer for virtualization due to their lightweight
and secure runtimes; however, dependencies in docker are handled
through Linux distribution archives, which would be lacking in
the environment of a unikernel.  Without such tools, the art of
manually packing a static archive for linking or building each
shared library with the correct version of Glibc will remain out
of reach for all but the most skilled devops engineers deploying
in the most demanding situations where significant cost and
performance benefits of unikernels may offset the additional
development work required for deploying the unikernel. The build
tools tested during this study, unik and ops, were both unable to
contend with library dependencies in an efficient manner.

Beyond the deployment itself, there are supplementary
considerations that must be investigated in terms of the
performance implications of unikernels. Docker’s AUFS allows for
something which unikernel images, in their current,
statically-linked form, do not—deduplication of layers. For
example, if the unikernel is being used for a microservice-based
web API, it would not be uncommon for there to be two endpoints
that look very similar from a dependency point of view—endpoints
involved with creating and updating a user’s profile, for
one—which would duplicate all library code in each binary. In a
Docker deployment, the libraries for the operating system would
be shared on disk, as the containers are stored as layers and
extended with each command executed in the Dockerfile. This
benefit extends to memory, as well—different docker containers
descending from the same parent layers are able to share the same
pages in memory due to Kernel Samepage
Merging~\cite{dockerdedup},~\cite{ksm}.  Unikernels, on the other
hand, may be able to share less memory due to differences in how
private pages may be accessed by unikernels sharing a majority of
code, but being compiled with different static dependencies. This
is an area requiring further research to experimentally determine
the extent of the memory saving, and the concept of copy-on-write
deduplication of memory pages is currently subject to security
concerns discovered along with side channel
attacks~\cite{ksm}~\cite{sidechannelattacks}.

\section{Conclusion}

Unikernels present compelling benefits in terms of performance
and security for deploying applications to the fog or the cloud,
but currently face issues in regards to managing dependencies,
updates, and compatibility with 3rd party libraries. A solution à
la \textit{docker build} for unikernels—providing a method for
dependency management as well as possibly for sharing and
extending images others have made—may provide a more secure and
performant platform for future cloud computing needs.

\vspace{12pt}

\bibliographystyle{IEEEtran}

\bibliography{Leon-the-dark-side-of-unikernels}

\end{document}